\documentclass[11pt, letter]{article}
\usepackage[utf8]{inputenc} 
\usepackage{amsmath, amssymb, graphicx, cite, color, enumerate, graphicx, multirow, algpseudocode, algorithm}
\usepackage[margin=1in]{geometry}

\makeatletter
\renewcommand{\@biblabel}[1]{\quad#1.}
\makeatother

\date{}

\begin{document}


\title{\textbf{Ensemble Analysis of Adaptive Compressed Genome Sequencing Strategies} 
}

\author{Zeinab Taghavi 
\\
Department of Computer Science, Wayne State University, Detroit, MI
\\
ztaghavi@wayne.edu}

\maketitle

\begin{abstract} 

Acquiring genomes at single-cell resolution has many applications such as in the study of microbiota. However, deep sequencing and assembly of all of millions of cells in a sample is prohibitively costly. A property that can come to rescue is that deep sequencing of every cell should not be necessary to capture all distinct genomes, as the majority of cells are biological replicates. Biologically important samples are often sparse in that sense. In this paper, we propose an adaptive compressed method, also known as distilled sensing, to capture all distinct genomes in a sparse microbial community with reduced sequencing effort. As opposed to group testing in which the number of distinct events is often constant and sparsity is equivalent to rarity of an event, sparsity in our case means scarcity of distinct events in comparison to the data size. Previously, we introduced the problem and proposed a distilled sensing solution based on the breadth first search strategy. We simulated the whole process which constrained our ability to study the behavior of the algorithm for the entire ensemble due to its computational intensity.

In this paper, we modify our previous breadth first search strategy and introduce the depth first search strategy. Instead of simulating the entire process, which is intractable for a large number of experiments, we provide a dynamic programming algorithm to analyze the behavior of the method for the entire ensemble. The ensemble analysis algorithm recursively calculates the probability of capturing every distinct genome and also the expected total sequenced nucleotides for a given population profile. Our results suggest that the expected total sequenced nucleotides grows proportional to $\log$ of the number of cells and proportional linearly with the number of distinct genomes. The probability of missing a genome depends on its abundance and the ratio of its size over the maximum genome size in the sample. The modified resource allocation method accommodates a parameter to control that probability.

\noindent {\bf Availability:} 
The squeezambler 2.0 C++ source code is available at \\{\tt http://sourceforge.net/projects/hyda/}.\\
\noindent The ensemble analysis MATLAB code is available at \\ {\tt https://sourceforge.net/projects/distilled-sequencing/}.

\end{abstract}


\section{Introduction}
Progress in DNA amplification techniques \cite{Chitsaz11} and high throughput cell cultivation methods \cite{Zengler02, Fitzsimons13} allow capturing of genomes at single-cell resolution. However, deep sequencing and assembly of all of the cells in a sample is prohibitively costly since there are millions sometimes billions of cells. The good news is that, to capture all distinct genomes, deep sequencing of every cell should not be necessary as the majority of cells are biological replicates. For instance, the number of detected distinct species in the human gut was estimated to be in the order of 1,000, while the number of microbial cells in a human body, most of which reside in the gut, is in the order of 100 trillion \cite{Qin10}. We call this effect the {\it sparsity} of distinct genomes in a sizeable microbial population. Biologically important samples are often sparse in that sense. We use sparsity to capture all of the genomes in a sample. 

During the last decade, the rich field of compressed sensing in non-adaptive \cite{Candes05, Candes06, Donoho06} and adaptive (distilled sensing and refinements) forms \cite{Haupt11,  Wei12} has been developed to reduce the cost of sampling and reconstruction of sparse signals \cite{Erlich10, Stobbe12}.  In the general form of the problem, both adaptive and non-adaptive methods reduce the number of sensing in comparison to the non-adaptive na\"ive sensing, and make it proportional to the number of distinct events times log of the data size \cite{Malloy12}. However while adaptive compressed methods may seem more cumbersome than their non-adaptive counterparts, adaptive methods often improve detection and estimation performance \cite{Malloy12}.

Our problem is an instance of a larger class of problems called \emph{element distinctness}, which is a popular problem in massive data analysis with numerous applications and different variants including (i) finding if there are duplicates in a list, (ii) calculating the number of distinct elements (support size), and (iii) estimating the distribution of distinct element populations \cite{Kane10}. In the vast majority of element distinctness problems, the complexity of deciding if two elements are identical is of $O(\text{element size})$. We distinguish between different classes of the problem based on the size of an element in comparison with the size of the entire population.

In some of the problems such as estimation of the number of distinct words an author knows (e.g. Shakespeare) \cite{Efron75}, the size of an element is very small in comparison with the size of the problem.  In some others, hash functions are used to reduce the element size. 
Such variants of the problem have been investigated deeply to find optimal algorithms both in time and space. For data stream analysis, if $n$ is the size of the language of elements, then the space complexity of optimal probabilistic $(1\pm\epsilon)$-approximation algorithm is $O(\epsilon^{-2} + \log n)$ and its time complexity is $O(n)$ \cite{Kane10}. In such algorithms, each element is either completely sensed or not sensed at all, i.e., no partial sensing of an element. However, there are important variants of the problem in which each element is complex, such as the case of whole genome sequences. The contributions of this paper and its predecessor \cite{Taghavi13} is (i) introduction of this other class of problems with large element size with respect to the sample size, and (ii) the first adaptive compressed method to the best of our knowledge to solve an instance of this problem in the form of finding all distinct genomes with reduced sequencing effort in a sparse microbial sample.

Assume that the distinction of two cells is based on the differences between their genomes. Therefore, the complexity of pairwise distinction is a function of the lengths of the DNA sequences, each in the order of $1,000,000-10,000,000$ base-pairs for a bacterial cell and $3,300,000,000$ base-pairs for a human cell, with an average size of $m$. A sample contains $n$ cells, for instance $10,000,000$ cells, where $n$ and $m$ are in the same order.  In this problem, there are two types of cost: (i) wet-lab cost related to sequencing, i.e., reading the DNA sequence digitally, and (ii) computational cost of genome assembly and comparison, i.e., digital reconstruction of the whole genome sequences from the sequencer output. The output reads of a sequencer are short randomly sampled subsequences of the genomic sequence which cover the genome multiple times. The number of reads that contain a genomic location is called the coverage. In the assembly, the reads are concatenated to reconstruct the whole genome. Sequence assembly is a challenging task due to sequencing errors and repetitive elements. To compare a number of sequenced read data sets, a co-assembly software such as HyDA \cite{Movahedi12} is used. The output of HyDA provides us with measures to compare the extent of similarity between the underlying genomes from which the read data sets are derived.

Wet-lab cost includes the monetary cost which is linearly proportional to the total number of base-pairs sequenced. If $m$ is the average genome size and $c$ is the necessary coverage, then the cost is $O(nmc)$ for the exhaustive sequencing of all cells. Computational cost includes the space complexity and time complexity of assembly and comparison. If the assembly is done using the de Bruijn graph \cite{Compeau11}, the time complexity is $O(nmc\log m)$ and space complexity is $O(nm)$. For instance, a typical real-world scenario involves $c=20$, $m=5,000,000$ bps, and $n=10,000,000$ for which the exhaustive wet-lab, time, and space cost complexities would be respectively $O(10^{15})$, $O(10^{16})$, and $O(10^{13})$. The exhaustive approach is not tractable even for a small population. Hence, sublinear algorithms are needed to solve the problem.

We propose an adaptive compressed method, also known as distilled sensing. Our ultimate goal is to reduce the $nm$ factor in wet-lab, time, and space complexities to $s m\log n$ in which $s$ is the number of distinct genomes in the community. We cannot use the algorithms and analyses given for the classical compressive sensing approach since our sparsity is \emph{unordered} set sparsity rather than ordered sparsity by time or space. As opposed to group testing in which the number of distinct events is often constant and sparsity is equivalent to rarity of an event, sparsity in our case means scarcity of distinct events in comparison to the data size. It is also important to note that we do not have positional access (a.k.a. random access in the computer science literature) to the DNA sequence, which limits the use of many dimensionality reduction techniques \cite{Johnson84}.

We previously defined the problem and proposed a distilled sensing solution based on the breadth first search strategy \cite{Taghavi13}. To evaluate the performance of our algorithm, we simulated the whole process including genome amplification by MDA \cite{Taghavi12}, sequencing by Illumina, (co-)assembly by HyDA \cite{Movahedi12}, and comparison. We proposed an adaptive resource allocation method to determine the amount of sampling of each genome in each round, which is related to the one proposed by \cite{Haupt11, Wei12}. Due to the computational intensity of each of those processes, we were able to demonstrate the power of our approach for a few instances of the problem, but the behavior of the algorithm for the entire ensemble is yet to be studied.

In this paper, we give a new algorithm based on the depth first search strategy and modify our previous breadth first search resource allocation and set selection. Since simulating the entire process is time-consuming, we provide a dynamic programming algorithm to analyze the behavior of the method over the entire ensemble. Our algorithm recursively calculates the probability of capturing every distinct genome and also the expected total sequenced nucleotides for a given population profile. It is important to note that even though the population is known for the ensemble analysis algorithm, the actual sensing algorithm works without that knowledge. That is our sensing algorithm can be applied to any population, even without knowing the profile. To have a clear view of the effect of each parameter on the expected cost, we assume that our model is error free at this stage. The results in this paper may lead to theoretical solutions and analysis with more complete model assumptions in the near future.

\section{Method}
Our method consists of two parts: (i) wet-lab process, and (ii) computational process. On the wet-lab side, we are assuming to have a high throughput device which is capable of isolating each cell in the sample, cultivate it, then extract the DNAs of each cultivated cell and amplify them. This device should also be capable of sampling customized amount from selected amplified DNAs, pool them, and prepare them for sequencing. If we would like to sequence more than one pool of samples in the same run, the device should uniquely barcode each pool before sending the samples for sequencing. Although there is currently no such device, one can envision automated microfluidic devices in near future based on the technologies already developed for separation, cultivation, DNA extraction, amplification, and barcoding \cite{Zengler02,Fitzsimons13}.

The output of sequencing is a library of reads which will be demultiplexed based on the barcodes. Therefore, for each pool of sampled amplicons which is sent for sequencing, a read data set is obtained. All the read data sets at each round are co-assembled (with {\tt HyDA} \cite{Movahedi12}). In the co-assembler, to each read data set a unique color is assigned. All the colors are assembled on a single de Bruijn graph. The output is a list of contigs and their colored average coverages. This provides us with a measure to compare the similarities between the assembly of different colors. Based on those similarities, we decide if any two assemblies could potentially be from the same genome.    

In the na\"ive exhaustive approach, each isolated cell is sampled and deeply sequenced. Based on the similarity measures provided by the co-assembler,  distinct genomes are then identified. In the adaptive method, at each round a number of collections of cells are selected. For each collection, the amount to be sampled from each cell is computed based on the analysis in the previous rounds. The output read data sets are analysed and the next round of sampling is calculated. We describe the details of the sampling collections and size in this section. First, let's clarify the assumptions for our model.

\subsection{Model assumptions}
The definition of distinct genomes may vary in different applications. We, instead of phenotypic notions like species or strains, use a quantifiable genomic measure to determine the distinction of genomes. We define two genome sequences to be \emph{distinct} if the ratio of their differences over the whole genome size is above a threshold, called $\tau$. That threshold is input by the user and controls a trade-off between sensitivity and specificity \cite{Taghavi13}. 

Let $C = \{C_1, C_2, \ldots, C_n\}$ comprise the input community of cells. As described earlier, we are given a device that can sense each cell $C_i$ partially at random, and the cost of a sensing is proportional to the sensing size, i.e., the number of nucleotides sampled. As the sensing size increases, the reconstructed genome of $C_i$ after the assembly converges to completion. To introduce appropriate notations, let $I \subseteq C$ be a subset of the community. Let $A$ be the sensing of the aggregated cells in $I$, which is the superposition of all sensing taken from the cells in $I$, i.e., the aggregated read data set or equivalently the resulting assembly. The key observation is that if there are enough replicates of a particular distinct genome in $I$, then that distinct genome can be completely captured from the superposition of partial sensing of the replicates provided that the partial sensing are random and unbiased. 

\subsection{Comparison of assemblies of two sets}\label{sec:comp}

Let $I_1, I_2 \subseteq C$ be two subcollections of the input community, and $A_1$, $A_2$ the corresponding aggregate sensing. If all of the distinct cells represented in $I_1$ are also represented in $I_2$, then we say that $A_1$ is \emph{subsumed} by $A_2$ $(A_1 \preceq A_2)$. In ideal world with no errors and genome variations, $A_1$ is called subsumed in $A_2$ if $A_1$ is a subset of $A_2$. However, in real world, while two genomes are considered similar (from the same type), they may have some variations like single nucleotide polymorphisms. In addition, errors, noise, and contaminations in sequencing and assembly make the situation harder to handle just by pure mathematical subset definition. To address this issue, the subset definition is relaxed to ignore those differences between two assemblies that are less than a threshold, $\tau$. Therefore, subsumption \cite{Taghavi13} is defined as follows 
\begin{equation}
A_1 \preceq A_2 \mbox{ iff } 0 \leq D_\tau(A_1, A_2).
\label{equ:subsumed}
\end{equation}
In this equation, $D_\tau(A_1, A_2)$ quantifies the differences in assembly of $A_1$ with respect to $A_2$ which is more than $\tau$ and is defined as follows
\begin{equation}
\label{equ:D}
D_\tau(A_1,A_2) = \tau - \frac{\|A_{1} \backslash A_{2}\|}{\|A_{1}\|},
\end{equation}
in which $A_{1} \backslash A_{2} = \{b \in A_1 | b \notin A_2 \}$, $\|\cdot\|$ denotes the total assembly size. In other words, $\tau$ is the maximum differences tolerated between two genomes which are considered similar. Parameter $\tau$ is user defined and $\tau$ gives a trade-off between specificity and sensitivity of the algorithm to distinguish between two distinct genomes \cite{Taghavi13}.

\subsection{Search strategies}
Our algorithm aims to assemble all of the distinct genomes represented in $C$ and identify at least one cell per distinct genome. The objective is to minimize the total number of bases required to be sequenced. To reach this goal a search tree is created and explored iteratively to find the leaves which are the sequenced and assembled species. In the first iteration, the set of deeply sequenced and completely assembled distinct genomes, $\mathcal{I}$, and its aggregated sensing, $\mathcal{A}$, is empty. The algorithm divides the $n$ cells $C_1, \ldots, C_n$ into two sets  $I_{1,1}= \{C_1, \ldots, C_{\lfloor n/2 \rfloor} \}$ and $I_{1,2}= \{C_{\lfloor n/2 \rfloor+1}, \ldots, C_n \}$. Denote $\bar I^1=\{I_{1,1}, I_{1,2}\}$. In each iteration $i$, $I_{i,j}$'s are subsets of $C$ and are chosen based on the results in the iteration $i-1$. The search tree of $I_{i,j}$ to find leaves can be traversed by different methods. Here we choose two methods, breadth first search (BFS) and depth first search (DFS). In the BFS strategy, in each iteration $i$, all $I_{i,j}$'s are explored at the same time, while in DFS, nodes ($I_{i,j}$'s) are explored sequentially in time and analyzed one after the other. 
 
In the recursive call on $\bar I^i=\{I_{i,1}, \ldots, I_{i,m_i}\}$, the set of cells is sensed according to the resource allocation policy. Then, the aggregated sensing, $A_{i,j}$, for each $I_{i,j}$ is obtained by sequencing and assembly. Those $I_{i,j}$'s that contain a single cell, i.e., $|I_{i,j}| = 1$, are leaves, and if they are fully assembled, they will be added to the list of deeply sequenced cells. In other words, if the corresponding assembly is reliable, i.e., $c_{i,j} \geq M_l$, for a given constant $M_l$, $I_{i,j}$ will be popped from $\bar I^i$ and pushed to $\mathcal{I}$. In addition, $A_{i,j}$ will be added to $\mathcal{A}$. 

For the BFS search strategy to find the optimum path to continue, a subset of $\bar I^i$ with minimum number of cells is chosen that covers all of the assembly. In other words, the minimum assembly-set cover $\bar I_{cover} \subseteq \bar I^i$ with minimum number of cells is found for which $\bar A \cup \mathcal{A}$ is subsumed in $\bar A_{cover} \cup \mathcal{A}$, i.e.,

\begin{align}
D_\tau(\bar A \cup \mathcal{A}, \bar A_{cover} \cup \mathcal{A}) & =  \tau - \frac{\| ( \bar A \cup \mathcal{A} ) \backslash ( \bar A_{cover} \cup \mathcal{A} )  \|}{\| \bar A \cup \mathcal{A} \|} \nonumber \\
& =  \tau - \frac{\| \bar A \cup \mathcal{A} \| - \|\bar A_{cover} \cup \mathcal{A}\|}{\| \bar A \cup \mathcal{A} \|} \geq 0. \label{equ:cover}
\end{align}

Second line can be derived from the first line because $( \bar A_{cover} \cup \mathcal{A} ) \subseteq ( \bar A \cup \mathcal{A} ) $. In these notions $\bar A_{cover}$ and $\bar A$ are the the resulting superposition of partial sensing and equivalently the corresponding assemblies of all cells represented in $\bar I_{cover}$ and $\bar I^i$, respectively. The search of the subtrees rooted at $I_{i,j} \notin \bar I_{cover}$ are terminated, and the next level set $\bar I^{i+1}:= \bar I_{cover}$.

For the DFS strategy, the minimum set cover is calculated gradually during several iterations. Since in the DFS, in each iteration $\bar I^i$ only includes two subsets, and the number of cells in both subsets are (almost) equal, the minimum set cover can be calculated based on the greedy algorithm. The $I_{i,j}$ with maximum assembly size has the highest priority to be in the minimum set cover. Therefore, $\bar I_{cover} :=\{ I_{i,j} \}$, and the second $I_{i,j'}$ will be pushed to the stack $\bar W$, which is the waiting list of the untraversed nodes in the tree. If $A_{cover}$ is subsumed in $\mathcal{A}$, then $A_{cover}$ will be emptied and the last element will be popped from $\bar W$ and pushed to $\bar I_{cover}$. This will continue until $A_{cover}$ is not subsumed in $\mathcal{A}$. In the end, the next level set $\bar I^{i+1}:= \bar I_{cover}$.
 
\begin{figure}[h!]
\begin{center}
\includegraphics[width=0.5\textwidth]{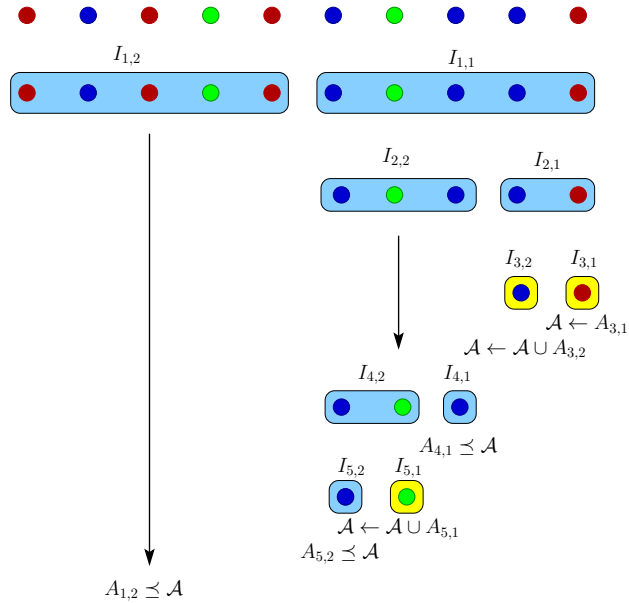}
\end{center}
\caption{{ \bf DFS algorithm example.} The adaptive depth first search algorithm for an example with 10 cells and 3 distinct genomes shown in different colors. Each row corresponds to one sequencing round. Yellow boxes represent leaves.}\label{fig:example}
\end{figure}

\begin{figure}[h!]
\begin{center}
\includegraphics[width=0.5\textwidth]{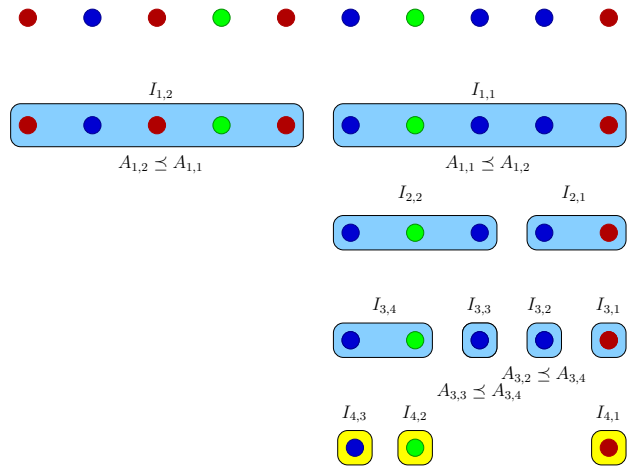}
\end{center}
\caption{{ \bf BFS algorithm example.} The adaptive breadth first search algorithm for an example with 10 cells and 3 distinct genomes shown in different colors. Each row corresponds to one sequencing round. Yellow boxes represent leaves.}\label{fig:BFSexample}
\end{figure}

For both search strategies, all subsets in $\bar I^{i+1}$ will be divided to two almost equal size subsets, which concludes iteration $i$. This algorithm will  continue until $\bar I^{i}$ and $\bar W$ are empty. Figures \ref{fig:example} and \ref{fig:BFSexample} depict examples of the DFS and BFS strategies on 10 cells with 3 distinct genomes shown in different colors.

\subsection{Resource allocation}

Resource allocation policy determines the size of partial sensing from each cell in each step. This is done with two objectives: (i) the amount of sensing from each element is such that with a given probability all of the distinct genomes present in $I_{i,j}$ can be reconstructed almost completely from the superposition of partial sensing, and (ii) the total sensing size in the whole algorithm is minimized.

Assume the input set of cells $I_{i,j}$ is obtained from splitting $I_{i-1,k}$ (for clarity we call it $I_{parent}$). Let $t_{parent}$, $a_{parent}$, and $c_{parent}$ be total nucleotides sampled, the assembly size, and the average coverage of $I_{parent}$ and $c'_{i,j}$ and $a'_{i,j}$ be the intended coverage and assembly size of $I_{i,j}$. We assume that there is a constant minimum coverage $M_u$, such that if the coverage is above $M_u$, then the resulting assembly covers the entire genome, i.e., does not have any gaps. We would like the actual coverage $c_{i,j}$ after the sequencing and assembly to be at least $M_u$, so we let $c'_{i,j} = M_u$ as a surrogate. Hence, the total nucleotides $t_{i,j}$ to be sampled and sequenced from $I_{i,j}$ is estimated by
\begin{equation}
\label{equ:t}
t_{i,j} = \frac{t_{parent} c'_{i,j} a'_{i,j}}{c_{parent} a_{parent}} = M_u \frac{t_{parent} a'_{i,j}}{c_{parent} a_{parent}}.
\end{equation}
In this equation $a'$ should be estimated from $c_{parent}$ and $a_{parent}$ and may differ from the actual value $a$ obtained after sequencing and assembly. If $I_{i,j}$ is a leaf, i.e., $|I_{i,j}| = 1$, then the algorithm does a deep sequencing of the single cell in $I_{i,j}$. In that case, the algorithm repeats the resource allocation and sequencing until a sufficient actual coverage is reached, i.e., $c_{i, j} \geq M_u$.

\subsection{DFS versus BFS}

Although the two search strategies are similar, they have differences in several aspects:
 \begin{itemize}
 \item{In DFS, the number of cell subcollections to be explored in each round is fixed. For instance, it is two in the current implementation. In BFS, this number is dynamic. For instance, in the first round of the example in Figure \ref{fig:BFSexample}, it is two and in the third round is four. From another point of view, the number of rounds in BFS is fixed, $\lceil \log_2 n \rceil$, whereas that in DFS is variable depending on the setup. The number of rounds and the number of subcollections are both desired to be minimum because each incurs a cost: each round incurs a setup cost for sample preparation and running-and-stopping the sequencer and each subcollection requires a unique barcode and incurs the cost of barcoding and sequencing the base pairs in the barcode. Hence, there is a trade-off between the two costs that determines the most suitable algorithm. The optimal algorithm should consider both costs.}
 \item{ In the process of choosing the minimum set cover, sets are compared to determine the subsumption relationships. In DFS, one side of the comparison is always $\mathcal{A}$ in which each single cell is deeply sequenced and completely assembled. Therefore, our information about that side is almost complete and has minimal error. The resource allocation of the algorithm is applied such that $A_{i,j}$'s for all $i$ and $j$ are completely sequenced on average to reduce unwanted missing of some distinct genomes. If the resulting coverage of $A_{i,j}$ is low, i.e., $c_{i,j} \leq M_l$ for a constant $M_l$, then the data is considered unreliable and $I_{i,j}$ is treated as if it contains a new distinct genome. In that case, $I_{i,j}$ would be divided into two groups and explored further in the following rounds with increased requested resources. Although $A_{i,j}$'s have deep coverage on average, there is a non-zero probability of missing small and less abundant distinct genomes. We explore these probabilities in the Results section. In BFS, the missing probability is more since on both sides of the $\preceq $ relation, cells are deeply sequenced not individually but on average, which increases the error and intrinsic noise in comparisons.} 
 \end{itemize}
The mentioned differences between the two strategies, do not result in considerable performance priority of one over the other method. Each are proper for a specific condition. In this paper, we compare the two methods using case-by-case setups. We show that the total nucleotides required is almost the same in both methods. We could not do the ensemble analysis for BFS using dynamic programming since the exploration of different subcollections in one round are coupled. For instance, the four subcollections to be compared in the third round of Figure \ref{fig:BFSexample} come from two parent subcollections in the second round. That creates an inevitable coupling between the parents in the second round. Without dynamic programming, exploring all of the permutations of cells to provide ensemble analysis is intractable. Therefore, we provide the ensemble analysis only for the DFS algorithm. 
 
\subsection{Implementation}

The pseudocode of the algorithm is given in \ref{alg:D&C}, \ref{alg:selectNextLevelSets}, and \ref{alg:subsumed}. \textsc{CompressedSearch} is the main function and \textsc{SelectNextLevelSets} and \textsc{Subsumed} are two subfunctions of the algorithm. This algorithm has been implemented in the tool {\tt squeezambler 2.0}.

\subsubsection{{\tt squeezambler 2.0} versus {\tt squeezambler 1.0}}
The tool {\tt squeezambler 1.0} has been implemented based on the BFS algorithm given in \cite{Taghavi12}. There are three main differences between the BFS algorithm implemented in  {\tt squeezambler 1.0} and the one in {\tt squeezambler 2.0}:
\begin{itemize}
 \item{In the recursion, the method to choose subsets passed to the next level is different in the two implementations. In {\tt squeezambler 1.0}, every subset that is subsumed in another one is eliminated from further analysis. However, this is not the optimum method to choose next level sets. In {\tt squeezambler 2.0},  a collection of subsets is chosen, which will cover the whole assembly with minimum number of cells.}
 \item{The resource allocation in {\tt squeezambler 1.0} was design for those sequencing technologies that have non-uniform coverage. That resource allocation results in assembly gaps and in some cases causes missing some of the distinct genomes. The resource allocation in {\tt squeezambler 2.0} is modified such that it reduces the random missing of the distinct genomes and let us predict the probability of missing genomes. This probability is analysed in the Results section.  We have added the resource allocation described in this paper to {\tt squeezambler 1.0}. The new version is called {\tt squeezambler 1.1}.}
 \item{The parameter $\tau$ in {\tt squeezambler 1.0} is variable and is dependent on the number of cells involved in each round. As the number of cells increases, $\tau$ decreases. Reduced $\tau$ increases the number of base pairs required to be sequenced when error appears in the reads. In {\tt squeezambler 2.0} $\tau$ is set to be fixed in the whole algorithm.}
\end{itemize}

\begin{algorithm}[h!]
\begin{algorithmic}[1]
\State {\bf Input:} $C = \{C_1, C_2, \ldots, C_n\}$
\State {\bf Output:} $\mathcal{A}$, $\mathcal{I}$
\State $I_{1,1} \gets \{C_1, \ldots, C_{\lfloor n/2 \rfloor} \} $ 
\State $I_{1,2} \gets \{C_{\lfloor n/2 \rfloor+1}, \ldots, C_n \}$ 
\State $\bar I \gets \{ I_{1,1}, I_{1,2} \} $ \Comment $\bar I$ is the list of the subsets to be analysed in the subsequent round
\State $\bar W \gets \{\}$ \Comment $\bar W$ is the waiting list of the subsets assembled but not ready to be analysed immediately
\State $i \gets 1$ \Comment $i$ is the sequencing round index
\While 	{\textsc{either $\bar I$ or $\bar W$ is not empty}}
	\State $\bar t \gets $ \textsc{ResourceAllocate}($\bar I$, $a_{parent}$, $c_{parent}$, $M_u$) \Comment $\bar t=\{t_{i,1},\ldots,t_{i,|\bar I|} \}$; based on Equ \ref{equ:t}
	\State $\bar A,  \bar a, \bar c \gets $ \textsc{SequenceAndAssemble}($\bar t$, $C$)  
	\Comment $\bar A = \{A_{i,1},\ldots,A_{i,|\bar I|} \}$; 
	$\bar c = \{c_{i,1},\ldots,c_{i,|\bar I|} \}$;  $A_{i,j}$, $c_{i,j}$ are the assembly set and the average coverage of $I_{i,j}$, respectively.
	\State $\bar I$, $\bar W$ $\gets$ \textsc{selectNextLevelSets($\bar I$, $\bar W$, $\bar A$,  $\bar c$, $F$)} \Comment $F$: DFS or BFS flag
	\State $i \gets i+1$
\EndWhile 
\end{algorithmic}
\caption{\textsc{Compressed Sequencing}}\label{alg:D&C}
\end{algorithm}

\begin{algorithm}[h!]
\begin{algorithmic}[1]
\State {\bf Input:} $\bar I$, $\bar W$, $\bar A$,  $\bar c$, $F$
\State {\bf Output:} $\bar I_{new}$, $\bar W_{new}$
\State $\bar L = \{\}$ \Comment list of subsets with low quality assemblies
\State $A_L = \{\}$ \Comment assemblies of subsets in $\bar L $
\For {$j = 1 \ldots |\bar I|$ }
	\If {$c_{i,j} < M_l$}
		\State \textsc{move $I_{i,j}$ from $\bar I$ to $\bar L$}  \Comment move all low coverage assembled  $I_{i,j}$ to $\bar L$
		\State \textsc{move $A_{i,j}$ from $\bar A$ to $ A_L$}  
	\Else	\If {$|I_{i,j}| = 1$}
				\State \textsc{move $I_{i,j}$ from $\bar I$ to $\mathcal{I}$}  \Comment move all single cell assembled $I_{i,j}$ to $\mathcal{I}$
				\State \textsc{move $A_{i,j}$ from $\bar A$ to $\mathcal{A}$}  
			\EndIf
	\EndIf 
\EndFor 
\If {$F$ is BFS}
	\State \textsc{find the minimum set cover $A_{new}$ corresponding to $\bar I_{new} \subseteq \bar I$ for which $D_\tau( (A \cup A_L \cup \mathcal{A}) , (\bar A_{new} \cup A_L \cup \mathcal{A}))\geq 0$ and $|\bar I_{new}|$ is minimum} \Comment  Equ. \ref{equ:cover}
	\State $\bar I_{new} \gets \bar I_{new} \cup \bar L$
	\State 	$\bar W_{new} = \{\}$
\Else \If {$F$ is DFS} 
	\State	$\bar W_{new} = \bar W$
	\State \textsc{pop all $I_{i,j}$'s from $\bar I_{new}$ and push to $\bar W_{new}$ except one subset with the maximum assembly size}
	\While {$|\bar I_{new}| = 0$ \textsc{and} $|\bar W_{new}| > 0$} 
		\State $I_{new} \gets $ \textsc{pop last subset in $\bar W_{new}$} 
		\State $A_{new} \gets $ \textsc{assembly of} $I_{new}$
		\If {\textsc{not Subsumed}($A_{new}$, $\mathcal{A}$)}  
			\State \textsc{push $I_{new}$ to $\bar I_{new}$}
		\EndIf
	\EndWhile
	\EndIf
\EndIf
\State \textsc{divide all $I_{i,j}$'s in $\bar I_{new}$ to two sets}
\end{algorithmic}
\caption{\textsc{selectNextLevelSets}}\label{alg:selectNextLevelSets}
\end{algorithm}

\begin{algorithm}[h!]
\begin{algorithmic}[1]
\State {\bf Input:} $A_{new}$, $\mathcal{A}$
\State {\bf Output:} $r \in \{\mbox{ true}, \mbox{ false }\}$
\State
\If {$c \leq M_l$} \Comment low quality assembly; explore the node further.
	\State $r \gets $ false
\Else
	\State $D \gets \tau - \| A_{new} \backslash \mathcal{A}\| / \|A_{new}\|$ \Comment Equ \ref{equ:D}
	\If {$D < 0$} \Comment Equ \ref{equ:subsumed}
		\State $r \gets $ false
	\Else
		\State $r \gets $ true
	\EndIf
\EndIf
\end{algorithmic}
\caption{\textsc{Subsumed}}\label{alg:subsumed}
\end{algorithm}

\subsection{Ensemble analysis}
We provide the ensemble analysis on the DFS algorithm to calculate the expected total number of sequenced base pairs and the probability of capturing every distinct genome over the entire ensemble of $n!$ permutations of $C$. To reduce the complexity, instead of exhaustively trying multiple permutations we developed a dynamic programming algorithm to calculate the results. In this analysis, the simulation of the entire process is replaced by black boxes which are  mathematical models of the behaviors of the process. To ease finding decoupled effects of different parameters in the algorithm, we do not consider the sequencing and assembly errors in our model. Another assumption, again in the interest of other important sparsity-related parameters, is to consider uniformity of coverage. This assumption is not far from reality. With the advancement of automated microfluidic cell separation and cultivation devices \cite{Zengler02,Fitzsimons13}, the genome can be captured from cultivated cells and sequenced with close to uniform coverage. This is different from the assumption we made in \cite{Taghavi13} for which a genome was amplified from a single cell using multiple displacement amplification and suffered from highly uneven coverage after sequencing. Although we are assuming uniform coverage distribution in this work as opposed to in \cite{Taghavi13}, this is only a convenient choice that does not change the algorithm. This assumption is reflected only in (\ref{equ:cov}).

Given the uniformity of coverage, we assume
\begin{equation}
a(c) = \left\{ \begin{array}{ll}
 g\dfrac{c}{M_u} & c \leq M_u,\\
g & c > M_u,
\end{array} \right.\label{equ:cov}
\end{equation}
in which $c$ is the sequencing coverage, $a$ is the total assembly size, $g$ is the genome size, and $M_u$ is a constant that defines the minimum coverage to obtain a complete assembly of the whole genome.  For more advanced models, see \cite{Motahari13}. In the ensemble analysis, we treat (co-)assembly as a black box oracle that knows $g$ and $M_u$, the input to which is the total sequenced nucleotides and the output of which are $a$ and $c$. That is based on the assumption that $ac$ is the total sequenced nucleotides, i.e. there are no sequencing errors. In this case the resource allocation formula in (\ref{equ:t}) will be reduced to $t_{i,j} = M_u a'_{i,j}$. In the worst case, all of the distinct genomes in $I_{parent}$ are also represented in $I_{i,j}$ which means $a'_{i,j} \geq a_{parent}$. Therefore, the total nucleotides can be estimated by 
\begin{equation}
\label{equ:t ensemble}
t_{i,j} =  2 M_u a_{parent}.
\end{equation}
which is twice the lower bound as a safe margin.

\subsubsection{Dynamic programming algorithm}
The dynamic programming algorithm can be divided into three main functions \textsc{Cost} (Algorithm \ref{alg:cost}), \textsc{AllocateSequenceAssembleOracle} (Algorithm \ref{alg:oracle}), and \textsc{Subsumed} (Algorithm \ref{alg:subsumedEnsemble}). Algorithm \ref{alg:cost} is the main dynamic programming, and its subroutines are presented in Algorithms \ref{alg:oracle} and \ref{alg:subsumed}. Let $s$ be the number of distinct genomes in $C$. A distinct genome profile $p = (p_1, p_2,\ldots, p_s) \in \left(\mathbb{N} \cup \{0\}\right)^s$ is a population vector. In the root of the search tree, $\sum_{j=1}^s p_j = n$, where $n$ is the total number of cells. The vector of deeply sequenced and assembled distinct genomes before exploration of the current node is denoted by $\mathcal{I} = (\mathcal{I}_1, \mathcal{I}_2, \ldots, \mathcal{I}_s)$ where $\mathcal{I}_j \in \{0, 1\}$ and $\mathcal{I}_j = 1$ means that the distinct genome $j$ has been sequenced and completely assembled. Throughout the algorithms, $\bar a = (a_1, a_2, \ldots, a_s)$ is the assembly size profile per distinct genome in the current node, and $\| \bar a\|_1$ is the total assembly size. Denote the assembly size of the parent search node by $a_{parent}$, total sequenced nucleotides by $t$, and the expected total sequenced nucleotides by $\mathbb{E}[t]$. We denote the probabilities of capturing distinct genomes by $\mathbb{P} = (q_1, q_2, \ldots, q_{2^s})$, in which $q_j$ is the probability of the vector of deeply sequenced and assembled distinct genomes $\mathcal{I} \in \{0, 1\}^s$ where $j = \langle \mathcal{I}, (2^0, 2^1, \ldots, 2^{s-1})\rangle + 1$, upon complete exploration of the current node in the search tree. Above, $j$ is the decimal representation of $\mathcal{I}$ treated as binary (in reverse order) plus one. For example, for $s = 3$ and $\mathcal{I} = (0,1,1)$ (in short $\mathcal{I}=011$), $j$ is the decimal representation of reverse of $011$ plus one which is equal to $7$. Therefore, $q_7$ (or in another notation $q_{011}$) is the probability that after exploration of the current node the distinct genomes 2 and 3 are recognized by the algorithm, while distinct genome 1 is missed. 

The \textsc{Cost} function requires the genome population profile for the current node, $p$, the set of distinct genomes already deeply sequenced and assembled, $\mathcal{I}$, and the result of the assembly of the parent node $a_{parent}$ as input parameters. The output of this function is the estimated cost $\mathbb{E}[t]$ and the probabilities of capturing genomes $\mathbb{P}$ after exploring the current node. At first (line 5), an oracle will decide on the sampling size of the current node and the resulting coverage and assembly size based on the formulations given in (\ref{equ:t}) and (\ref{equ:cov}). In line 7, using the function \textsc{Subsumed}, the node is then compared with $\mathcal{I}$ to see if it includes any new distinct genome. If there is no new distinct genome (line 41), the probability of capturing the corresponding genomes is set to 1 and the function exits. Otherwise, the node will be explored further. 

If the node includes only one cell, then that node is a leaf, and it will be sequenced and assembled deeply (line 9) and the corresponding capturing probability is set to 1. In the case of a node with more than one cell, the collection of cells will be divided into two groups. The expected cost and capturing probability, starting from the current node, is calculated over the ensemble of all of the possible divisions of the node into $v$ and $w$ between lines 19 and 39. The ensemble parameters are averaged over all divisions $(v,w)$ by calculating their probability of occurrence (line 36). For a given division, $v$ is explored followed by $w$. For each $u \in \{v, w\}$ (line 24) and each combination of already captured distinct genomes with non-zero probability (lines 26-28), the expected cost $t'$ and capturing probability profile $\mathbb{P}'''$ are recursively calculated using the \textsc{Cost} function (line 29).  These parameters are averaged over all non-zero probability profiles (lines 30, 31).

\section{Results}
\subsubsection{DFS versus BFS}
\begin{figure}[h!]
\begin{center}
\includegraphics[width=0.6\textwidth]{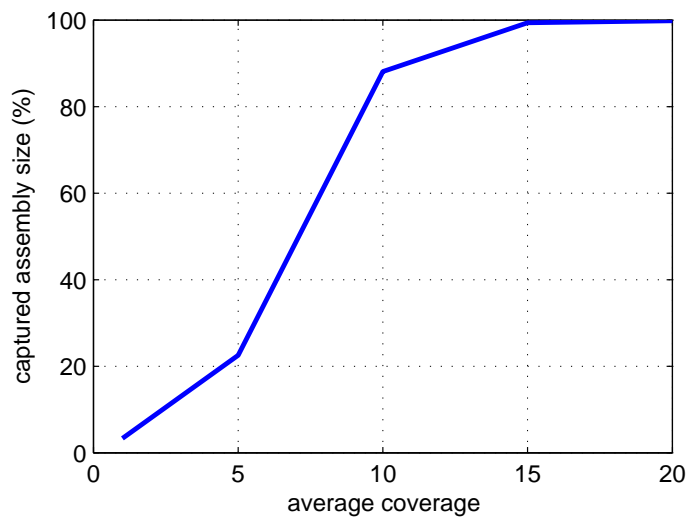}
\end{center}
\caption{Output assembly size in percentile versus coverage. The plot is the average of the output of sequencing simulation using ART \cite{Huang12} and assembly using HyDA \cite{Movahedi12} over 6 distinct genomes listed in Table \ref{tab:species}. }\label{fig:avsc}
\end{figure}

To compare the performance of DFS and BFS algorithms, we tested the algorithms on simulated data. We have selected 6 distinct genomes from human gut microbiome \cite{Qin10} to generate 3 setups (see Tables \ref{tab:species} and \ref{tab:experiments}). The genomes were amplified and sequenced using ART Illumina sequencing simulator \cite{Huang12}. Reads are 100 bp long, uniformly covering the whole genome. The assembler used in the paper is HyDA co-assembler \cite{Movahedi12}. To allocate resources, the relationship between the coverage and the assembly size of this setup is calculated using ART and HyDA over the 6 distinct genomes. The result is depicted in Figure \ref{fig:avsc}. We selected $M_u = M_l = 15$ and $\tau = 0.1$. 

We have compared the performance of four different methods: (i) na\"ive method of sequencing each cell deeply, (ii) BFS compressed method using {\tt squeezambler 1.1}, (iii) BFS compressed method using {\tt squeezambler 2.0}, and (iv) DFS compressed method using {\tt squeezambler 2.0}. The results are summarized in Table \ref{tab:results}. See that {\tt squeezambler 2.0} BFS outperforms {\tt squeezambler 1.1} BFS in total sequencing base pairs. The BFS and DFS algorithms of {\tt squeezambler 2.0} have close performances in terms of the total sequenced base pairs. However, for the case of 31 cells and 6 distinct genomes, BFS missed one  distinct genome. This is one of the examples that shows the genome missing probability is slightly more in BFS. Overall, both algorithms have comparable performances. Comparison of the performance of na\"ive algorithm and compressed methods shows that as the number of cells increases, the total sequenced base pairs increases linearly for the na\"ive algorithm and sublinear for the compressed methods. Although for the case of 31 cells, the na\"ive method outperforms the compressed method, at 140 cells the compressed method shows its strength.

\begin{table*}[h!]
\caption{6 distinct genomes (species) used in simulation.\label{tab:species}}
\begin{center}
{\begin{tabular}{|l|lcl|}
\hline
NCBI ID & Name & Ref. Status & Size (bps) \\\hline
NC\_004663.1 & Bacteroides thetaiotaomicron VPI-5482 chromosome & complete & 6.29 M \\
NC\_009614.1 & Bacteroides vulgatus ATCC 8482 chromosome & complete & 5.16 M \\
NC\_009615.1 & Parabacteroides distasonis ATCC 8503 chromosome & complete & 4.81 M \\
NC\_008532.1 & Streptococcus thermophilus LMD-9 & complete & 1.86 M \\
NC\_016776.1 & Bacteroides fragilis 638R & complete & 5.37 M \\
FP929051.1 & Ruminococcus bromii L2-63 & draft & 2.25 M \\
\hline
\end{tabular}
}{}
\end{center}
\end{table*}

\begin{table*}[h!]
\caption{Our simulation setups: (i) 31 cells; 6 distinct genomes, (ii) 59 cells; 4 distinct genomes, and (iii) 140 cells; 4 distinct genomes.\label{tab:experiments}}
\begin{center}
{\begin{tabular}{|l|cccccc|}\hline
\multirow{3}{*}{NCBI ID} & \multicolumn{6}{c|}{Abundance (\%)} \\\cline{2-7}
 & \multicolumn{2}{c}{31 cells;} & \multicolumn{2}{c}{59 cells;} & \multicolumn{2}{c|}{140 cells;} \\
 & \multicolumn{2}{c}{6 distinct genomes} & \multicolumn{2}{c}{4 distinct genomes} & \multicolumn{2}{c|}{4 distinct genomes} \\\hline
NC\_004663.1 & 11 & 35.5\% & 22 & 37\% & 35 & 25\%\\
NC\_009614.1 & 4 & 13\% & 7 & 12\% & 35 & 25\%\\
NC\_009615.1 & 3 & 10\% & 8 & 14\% & 35 & 25\%\\
NC\_008532.1 & 1 & 3\% & 0 & 0\% & 0 & 0\%\\
NC\_016776.1 & 1 & 3\% & 0 & 0\% & 0 & 0\%\\
FP929051.1 & 11 & 35.5\% & 22 & 37\% & 35 & 25\%\\
\hline
\end{tabular}}{}
\end{center}
\end{table*}

\begin{table*}[h!]
\caption{{\tt squeezambler} results for the three setups summarized in Table \ref{tab:experiments}. \label{tab:results}}
\begin{center}
{\begin{tabular}{|l|lcccc|}\hline
 		& 			& Total 		& 	 				&  					& \\
Setup	& Method  	& Sequencing 	& Max 	 			& No. of Predicted 	& Iterations\\
  		&  			& (Gbps) 		& Barcodes			& distinct genomes	&\\\hline
31 cells; & na\"ive &  4.0 & 31 & 6 & 1\\
6 distinct & squeezambler 1.1, BFS &  13.8 & 10 & 6 & 5\\
 genomes   & squeezambler 2.0, BFS &  11.2 & 8 & 5 & 5 \\
				   & squeezambler 2.0, DFS &  11.0 & 2 & 6 & 12\\\hline
59 cells; & na\"ive &  8.0 & 59 & 4 & 1\\
4 distinct & squeezambler 1.1, BFS &  10.0 & 4 & 4 & 6 \\
genomes & squeezambler 2.0, BFS &  8.7 & 4 & 4 & 6 \\
 & squeezambler 2.0, DFS &  8.8 & 2 & 4 & 9\\\hline
140 cells; & & & & & \\
4 distinct & na\"ive & 18.5 & 140 & 4 & 1\\
genomes & squeezambler 2.0, DFS & 10.3 & 2 & 4 & 10\\
\hline
\end{tabular}}{}\\
\end{center}
\end{table*}

\subsubsection{Ensemble analysis}
In this section, we selected $M_u = 5$, $M_l = 0.3$, and $\tau = 0.2$ except in Table \ref{tab:prob}. Bacterial genome sizes were considered to be within 1 -- 12 Mbps.
In the current version of the program implemented in MATLAB, we computed the results for a small number of cells and distinct genomes, i.e., $< 200$ cells and $< 11$ distinct genomes. However, we expect to be able to run the algorithm for a larger number of cells and distinct genomes with an advanced implementation in C++.
We tried to decouple the effect of different parameters in the analysis, namely $\tau$ in the algorithm, and the number of cells and species in the input. We would like to test whether the expected total cost complexity is $O(s)$ for a fixed $n$ and $O(\log n)$ for a fixed $s$ and population profile. This is a first step to show the expected total cost is $O(s \log n)$ in the future. Therefore, the results also provide intuition for a potential thorough theoretical analysis of the expected cost and capturing probability. 

\subsubsection{Expected cost}
\begin{figure}[h!]
\begin{center}
\includegraphics[width=0.8\textwidth]{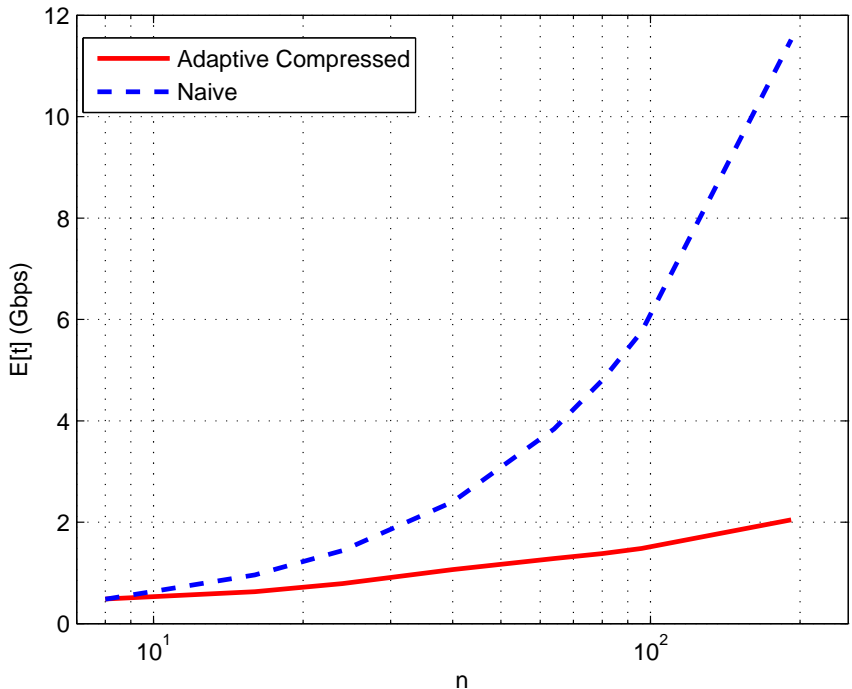}
\end{center}
\caption{$\mathbb{E}[t]$ for different number of cells $n$ in our adaptive compressed algorithm versus the na\"ive cell by cell sequencing. The total sequenced nucleotides in the  na\"ive case is $M_u \times \max$ genome size $\times n = 60 n$ Mbps. The population profile is $\frac{n}{8}(3, 2, 3)$ with genome sizes $(4, 12, 2)$ Mbps and $\tau = 0.2$. The genome of length 2 Mbps may not be captured with at most 3\% probability; the other two genomes are always captured. The $x$ axis is shown in log scale.}\label{fig:E-vs-n}
\end{figure}

To investigate the growth rate of $\mathbb{E}[t]$ for different number of cells $n$ and compare it with the na\"ive cell-by-cell sequencing, we ran our program for the profiles of $\frac{n}{8}(3, 2, 3)$ where $n=8,\ldots, 192$ is the total number of cells with genome sizes (4, 12, 2) Mbps. Figure \ref{fig:E-vs-n} depicts the results. The total sequenced nucleotides in the  na\"ive case is $M_u \times \max \mbox{genome size} \times n = 60 n$ Mbps. The genome of length 2 Mbps may not be captured with at most 3\% probability; the other two genomes are always captured. Figure \ref{fig:E-vs-n} suggests that $\mathbb{E}[t]$ grows almost linearly with $\log n$ whereas the na\"ive cost grows linearly with $n$. Hence, $\mathbb{E}[t] = O(\log n)$ for fixed number of distinct genomes $s$ and population profile.

\begin{figure}[h!]
\begin{center}
\includegraphics[width=0.8\textwidth]{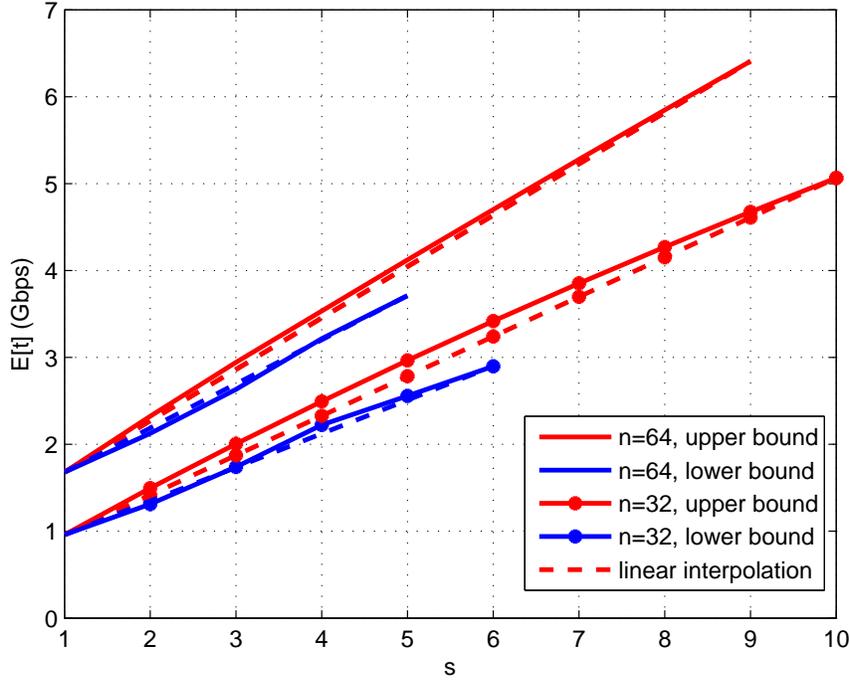}
\end{center}
\caption{$\mathbb{E}[t]$ for different number of distinct genome $s$ and $n=32, 64$ in our adaptive compressed algorithm. For each $n$, the best and worst population profiles in terms of expected cost are considered. The best case corresponds to roughly uniform $n/s$ cells per distinct genome and the worst corresponds to $n-s+1$ cells from one distinct genome and one cell per every other $s-1$ distinct genomes. To decouple the effect of genome sizes, the genome size is 4 Mbps for all distinct genomes and $\tau = 0.2$. Capturing probability of all distinct genomes in all cases is 1. Linear interpolated costs are plotted using dashed line for comparison.}\label{fig:E-vs-s}
\end{figure}

To characterize the behavior of $\mathbb{E}[t]$ for different number of distinct genomes $s$, we plotted $\mathbb{E}[t]$ versus $s=1,\ldots,10$ for $n=32, 64$ in Figure \ref{fig:E-vs-s}. For each $n$, the best and worst population profiles in terms of expected cost were considered. The best case corresponds to roughly uniform $n/s$ cells per distinct genome and the worst corresponds to $n-s+1$ cells from one distinct genome and one cell per every other $s-1$ distinct genomes (experimentally verified). The genome size was fixed at 4 Mbps for all distinct genomes to factor out the effect of genome sizes and $\tau = 0.2$. Capturing probability of all distinct genomes in all cases was 1. Comparing the experimental curves with the linearly interpolated cost curves in Figure \ref{fig:E-vs-s} suggests that the upper bound of cost (worst case) for each fixed $n$ is $O(s)$. In other words, the worst case cost grows almost linearly with the number of distinct genomes. Hence, for all population profiles,  $\mathbb{E}[t] = O(s)$ for fixed number of cells $n$. 

\subsubsection{Capturing probability}
We ran our program for a number of setups with 3 and 4 distinct genomes. The number of cells $n$ was fixed at 40 in all experiments. Genome sizes varied between 1 and 12 Mbps. The total na\"ive single-cell sequencing is the minimum coverage $M_u$ times the number of cells, which is 40, times the maximum genome size. In this case, the na\"ive total cost is $2.4$ Gbps. Tables \ref{tab:prof} and \ref{tab:prob} show the setups and their expected sequenced nucleotides $\mathbb{E}[t]$ and $\mathbb{P}$, the capturing probability. Recall $\mathbb{P} = (q_{00\cdots0}, q_{10\cdots0}, \ldots, q_{11\cdots1})$, in which $q_{\mathcal{I}}$ for $\mathcal{I} \in \{0, 1\}^s$ is the joint probability of capturing distinct genome $j$ if $\mathcal{I}_j = 1$ and missing it if $\mathcal{I}_j = 0$ for $j=1, 2, \ldots, s$.   

In Table \ref{tab:prof} with constant $\tau=0.2$, that genome whose length is 2 Mbps in the first row may not be captured with probability 0.62\% because $2/12 < 0.2=\tau$. We see a similar effect in the subsequent rows. Those genomes whose lengths are 1 and 2 Mbps may not be captured with non-zero probability. As the abundances of short genomes increase, the probabilities of missing them decrease. This suggests that those genomes whose lengths are shorter than $\tau$ times the largest genome size may not be captured with non-zero probability. 

To further investigate, we ran our program on other setups with varying $\tau$, which are presented in Table \ref{tab:prob}. For the profile (1, 1, 2, 4) and $\tau=0, 0.1$, all of the genomes are captured with probability 1. For $\tau=0.2$, there is a non-zero probability of missing the genomes of length 1 and 2 Mbps. For $\tau=0.4$, the genome of size 4 Mbps joins the other two short genomes, and the probability of missing it becomes non-zero. Comparing the profiles (1, 1, 2, 4) and (1, 1, 1, 5) for $\tau=0.4$, the probability of missing the genome of size 2 Mbps, i.e., $q_{1101}$, increases significantly from 0.8\% to 23\% as a result of decrease in its abundance, whereas the probability of not capturing the genome of size 1 Mbps, i.e., $q_{1110}$, decreases from 4.1\% to 0.8\% as its abundance increases. This suggests that the missing probability depends on abundance, potentially relative to the abundance of the largest genome in the population.

\begin{table}[h!]
\caption{Effect of population profile. In this setup $\tau = 0.2$ and the total number of cells is $n = 40$.  $q_{\mathcal{I}}$ for $\mathcal{I} \in \{0, 1\}^s$ is the joint probability of capturing distinct genome $j$ if $\mathcal{I}_j = 1$ and missing it if $\mathcal{I}_j = 0$ for $j=1, 2, \ldots, s$. Different values of $\mathcal{I}$ are explicitly given as binary strings in the fourth column. $\epsilon$ is a non-zero probability less than $10^{-4}$.}
\begin{center}
\begin{tabular}{|c|c|c|c|}
\hline
$p/5$ & Genome size  & $\mathbb{E}[t]$  & $\mathbb{P = }$\\
 & (Mbps) & (Gbps) & $(q_{000}, q_{100}, q_{010}, q_{110}, q_{001}, q_{101}, q_{011}, q_{111})$ \\
\hline
$(3, 2, 3)$ & $(4, 12, 2)$ & $1.066$ & $(0, 0, 0, 0.0062, 0, 0, 0, 0.9938)$ \\
\hline
 & & & $\mathbb{P = } (q_{0000}, q_{1000}, q_{0100}, q_{1100}, q_{0010}, q_{1010}, q_{0110}, q_{1110},$ \\
 & & & $q_{0001}, q_{1001}, q_{0101}, q_{1101}, q_{0011}, q_{1011}, q_{0111}, q_{1111})$\\
\hline
$(1, 1, 1, 5)$ & $(4, 12, 2, 1)$ & $1.383$ & $(0, 0, 0, 0, 0, 0, 0, 0.0020, 0, 0, 0, 0.0271, 0, 0, 0, 0.9709)$\\
\hline
$(1, 1, 2, 4)$ & $(4, 12, 2, 1)$ & $1.407$ & $(0, 0, 0, 0, 0, 0, 0, 0.0123, 0, 0, 0, 0.0001, 0, 0, 0, 0.9876)$\\
\hline 
$(1, 2, 3, 2)$ & $(4, 12, 2, 1)$ & $1.535$ & $(0, 0, 0, \epsilon, 0, 0, 0, 0.1687, 0, 0, 0, 0.0135, 0, 0, 0, 0.8179)$\\
\hline
\end{tabular}
\label{tab:prof}
\end{center}
\end{table}

\begin{table}[h!]
\caption{Effect of threshold $\tau$. The genome sizes are (4, 12, 2, 1) Mbps and the total number of cells is $n = 40$. $q_{\mathcal{I}}$ for $\mathcal{I} \in \{0, 1\}^s$ is the joint probability of capturing distinct genome $j$ if $\mathcal{I}_j = 1$ and missing it if $\mathcal{I}_j = 0$ for $j=1, 2, \ldots, s$. Different values of $\mathcal{I}$ are explicitly given as binary strings in the fourth column. $\epsilon$ is a non-zero probability less than $10^{-4}$.}
\begin{center}
\begin{tabular}{|c|c|c|c|}
\hline
$p/5$ & $\tau$ &$\mathbb{E}[t]$  &  $\mathbb{P} = (q_{0000}, q_{1000}, q_{0100}, q_{1100}, q_{0010}, q_{1010}, q_{0110}, q_{1110},$\\
 & & (Gbps) &  $q_{0001}, q_{1001}, q_{0101}, q_{1101}, q_{0011}, q_{1011}, q_{0111}, q_{1111})$\\
\hline
$(1, 1, 2, 4)$ & $0$ & $1.428$ & $(0, 0, 0, 0, 0, 0, 0, 0, 0, 0, 0, 0, 0, 0, 0, 1)$\\
\hline 
$(1, 1, 2, 4)$ & $0.1$ & $1.423$ & $(0, 0, 0, 0, 0, 0, 0, 0, 0, 0, 0, 0, 0, 0, 0, 1)$\\
\hline 
$(1, 1, 2, 4)$ & $0.2$ & $1.407$ & $(0, 0, 0, 0, 0, 0, 0, 0.0123, 0, 0, 0, 0.0001, 0, 0, 0, 0.9876)$\\
\hline 
$(1, 1, 2, 4)$ & $0.4$ & $1.266$ & $(0, 0, 0, 0.0002, 0, 0, 0.0008, 0.0411, 0, 0, \epsilon, 0.0794, 0, 0, 0.1621, 0.7165) $\\
\hline 
$(1, 1, 1, 5)$ & $0$ & $1.418$ & $(0, 0, 0, 0, 0, 0, 0, 0, 0, 0, 0, 0, 0, 0, 0, 1)$\\
\hline
$(1, 1, 1, 5)$ & $0.1$ & $1.414$ & $(0, 0, 0, 0, 0, 0, 0, 0, 0, 0, 0, 0, 0, 0, 0, 1)$\\
\hline
$(1, 1, 1, 5)$ & $0.2$ & $1.383$ & $(0, 0, 0, 0, 0, 0, 0, 0.0020, 0, 0, 0, 0.0271, 0, 0, 0, 0.9709)$\\
\hline
$(1, 1, 1, 5)$ & $0.4$ & $1.214$ & $(0, 0, 0, 0.0013, 0, 0, \epsilon, 0.0078, 0, 0, \epsilon, 0.2308, 0, 0, 0.1369, 0.6231)$\\
\hline
\end{tabular}
\end{center}
\label{tab:prob}
\end{table}

\begin{algorithm}[h!]
\begin{algorithmic}[1]
\State {\bf Input:} $p$, $\mathcal{I}$, $a_{parent}$
\State {\bf Output:} $\mathbb{E}[t]$ and $\mathbb{P}$
\State
\State $\mathbb{P}=(q_1, q_2, \ldots, q_{2^s}) \gets \mathbf{0}$ 
\State $t, \bar a, c \gets $  \textsc{AllocateSequenceAssembleOracle}($p$, $a_{parent}$) \Comment $t$ total nt, $\bar a = (a_1, \ldots, a_s)$, $c$ coverage
\State $\mathbb{E}[t] \gets t$
\If {not \textsc{SubsumedEnsemble}($\bar a$, $c$, $\mathcal{I}$)}
	\If {$\|p\|_1 = 1$} \Comment{leaf base case}
		\While 	{$c < 2M_u$} \Comment ensures the complete assembly for leaves
			\State $t, \bar a, c \gets $ \textsc{AllocateSequenceAssembleOracle}($p$, $\|\bar a\|_1$) 
			\State $\mathbb{E}[t] \gets \mathbb{E}[t] + t$
		\EndWhile 
		\State  $k \gets \arg\max p_j$
		\State  $\mathcal{I}^{new} \gets \mathcal{I}$
	    \State  $\mathcal{I}^{new}_k \gets 1$
	    \State  $j \gets \langle \mathcal{I}^{new}, (2^0, 2^1, \ldots, 2^{s-1}) \rangle + 1$
	    \State  $q_j = 1$ \Comment updating $\mathbb{P}$
	\Else	\Comment{recursion}
		\For {$v + w = p$, $v, w \in \left(\mathbb{N} \cup \{0\}\right)^s$}
			\State $t \gets 0$
			\State $\mathbb{P}' \gets \mathbb{P}$ \Comment $\mathbb{P}'=(q'_1, q'_2, \ldots, q'_{2^s})$
		    \State $j \gets \langle \mathcal{I}, (2^0, 2^1, \ldots, 2^{s-1}) \rangle + 1$
			\State $q'_j \gets 1$
			\For {$u \in \{v, w\}$} \Comment $v$ is explored followed by $w$
				\State $\mathbb{P}'' \gets \mathbf{0}$
				\For {$b$ \textsc{a binary vector of length} $s$}
				    \State $j \gets \langle b, (2^0, 2^1, \ldots, 2^{s-1}) \rangle + 1$
				    \If {$q'_j > 0$} \Comment average over all already captured distinct genome profile with non-zero probability
                        \State $t', \mathbb{P}''' \gets$ \textsc{Cost}($u$, $b$, $\|\bar a\|_1$)
                        \State $t \gets t + t' q'_j$
						\State $\mathbb{P}'' \gets \mathbb{P}'' + q'_j\mathbb{P}'''$
					\EndIf
				\EndFor
				\State $\mathbb{P}' \gets \mathbb{P}''$ \Comment for $w$, $q'$'s are updated
			\EndFor
			\State $\pi \gets  \prod_{j=1}^s {p_j \choose v_j}/{\|p\|_1 \choose \|v\|_1}$\Comment probability of $(v, w)$
			\State $\mathbb{E}[t] \gets \mathbb{E}[t] + \pi t$ \Comment average over all possible $(v, w)$
			\State $\mathbb{P} \gets \mathbb{P} + \pi\mathbb{P}'$ \Comment average over all possible $(v, w)$
		\EndFor
	\EndIf
\Else	\Comment{ all distinct genomes represented in $p$ have already been sequenced }
    \State  $j \gets \langle \mathcal{I}, (2^0, 2^1, \ldots, 2^{s-1}) \rangle + 1$
    \State  $q_j = 1$ \Comment updating $\mathbb{P}$
\EndIf

\end{algorithmic}
\caption{\textsc{Cost} - ensemble analysis main function}\label{alg:cost}
\end{algorithm}

\begin{algorithm}[h!]
\begin{algorithmic}[1]
\State {\bf Input:} $p$, $a_{parent}$
\State {\bf Output:} $t$, $\bar a$, and $c$
\State 
\State $t \gets 2M_u a_{parent}$ \Comment Equ \ref{equ:t ensemble}
\State $t_{pc} \gets t / \|p\|_1$ \Comment total sequenced nt per cell 
\State $\bar t_{pdg} \gets t_{pc}\cdot p$ \Comment total sequenced nt per distinct genome 
\State $\bar a, c \gets $ \textsc{Oracle}($\bar t_{pdg}$) \Comment oracle decides on the assembly size and coverage based on Equ \ref{equ:cov}
\end{algorithmic}
\caption{\textsc{AllocateSequenceAssembleOracle} - ensemble analysis}\label{alg:oracle}
\end{algorithm}

\begin{algorithm}[h!]
\begin{algorithmic}[1]
\State {\bf Input:} $\bar a$, $c$, $\mathcal{I}$
\State {\bf Output:} $r \in \{\mbox{ true}, \mbox{ false }\}$
\State
\If {$c \leq M_l$} \Comment low quality assembly; explore the node further.
	\State $r \gets $ false
\Else
	\State $x \gets \langle \neg \mathcal{I}, \bar a \rangle$ \Comment exclusive part of assemblies, $\neg$ is bitwise not, based on Equ \ref{equ:D}
	\If {$\tau - x / \|\bar a\|_1 < 0 $} \Comment Equ \ref{equ:subsumed}
		\State $r \gets $ false
	\Else
		\State $r \gets $ true
	\EndIf
\EndIf
\end{algorithmic}
\caption{\textsc{SubsumedEnsemble} - ensemble analysis}\label{alg:subsumedEnsemble}
\end{algorithm}

\subsection{Conclusion}
We presented an adaptive compressed algorithm for sequencing and \emph{de novo} assembly of distinct genomes in a bacterial community. We used the characteristics of sparsity of distinct genomes in a community of cells to decrease the amount of nucleotides needed to be sequenced. Using a dynamic programming algorithm to analyze the ensemble behavior of the algorithm, we showed that the expected cost is $O(\log(\text{number of cells in the community}))$ for fixed genome population profiles and $O(\text{number of distinct genomes})$ for fixed number of cells. Furthermore, we showed that for a non-zero threshold $\tau$, those genomes whose sizes relative to the maximum genome size in the community is less than $\tau$ may go undetected with a non-zero probability. This probability depends on the abundance of the corresponding genome. These results shed light on our future path towards theoretical analysis of our algorithm and further tree exploration strategies.

\bibliographystyle{unsrt} 
\bibliography{pub,ref}      

\end{document}